\documentclass{aa}
\usepackage{natbib}
\usepackage{graphicx}
\usepackage{epstopdf}
\usepackage{amsmath}
\usepackage{times}
\usepackage{url}
\usepackage{array}
\usepackage{float}
\usepackage{pgf}
\usepackage{pgfplots}
\usepackage{tikz}
\pgfplotsset{compat=1.9}
\usepackage{savesym}
\usepackage{amsmath}
\savesymbol{iint}
\savesymbol{iiint}
\usepackage{txfonts}
\restoresymbol{TXF}{iint}
\restoresymbol{TXF}{iiint}

\bibpunct{(}{)}{;}{a}{}{,} 

\newcolumntype{L}[1]{>{\raggedright\let\newline\\\arraybackslash\hspace{0pt}}m{#1}}
\newcolumntype{C}[1]{>{\centering\let\newline\\\arraybackslash\hspace{0pt}}m{#1}}
\newcolumntype{R}[1]{>{\raggedleft\let\newline\\\arraybackslash\hspace{0pt}}m{#1}}

\begin{document}

\title{A method of immediate detection of objects with a near-zero apparent motion in series of CCD-frames}

\author{Savanevych, V. E.\inst{1,2}
  \and Khlamov, S. V.\inst{2}
  \and Vavilova, I. B.\inst{3}
  \and Briukhovetskyi, A. B.\inst{1}
  \and Pohorelov, A. V.\inst{4}
  \and Mkrtichian, D. E.\inst{5}
  \and Kudak, V. I.\inst{2,6}
  \and Pakuliak, L. K.\inst{3}
  \and Dikov, E. N.\inst{7}
  \and Melnik, R. G.\inst{1}
  \and Vlasenko, V. P.\inst{1}
  \and Reichart, D. E. \inst{8}} 

\offprints{{Savanevych V. E., \email{vadym@savanevych.com}, Khlamov S. V., \email{sergii.khlamov@gmail.com}}}

\institute{Western Radio Technical Surveillance Center, State Space Agency of Ukraine, Kosmonavtiv Street, Mukachevo UA-89600, Ukraine
  \and Uzhhorod National University, Laboratory of space research, 2a Daleka Street, Uzhhorod UA-88000, Ukraine 
  \and Main Astronomical Observatory of the NAS of Ukraine, 27 Akademika Zabolotnogo Street, Kyiv UA-03143, Ukraine 
  \and Kharkiv National University of Radio Electronics, 14 Nauki Avenue, Kharkiv UA-61166, Ukraine 
  \and National Astronomical Research Institute of Thailand, 260 Moo 4, T. Donkaew, A. Maerim, Chiangmai TH-50180, Thailand
  \and Institute of Physics, Faculty of Natural Sciences, University of P. J. Safarik, Park Angelinum 9, Kosice SK-04001, Slovakia
  \and Scientific Research, Design and Technology Institute of Micrographs, 1/60 Akademika Pidgornogo Street, Kharkiv UA-61046, Ukraine
  \and Department of Physics and Astronomy, University of North Carolina at Chapel Hill, Chapel Hill NC-27599, North Carolina, USA} 

\date{Received 22 December 2016 / Accepted 25 September 2017}

\abstract{The paper deals with a computational method for detection of the solar system minor bodies (SSOs), whose inter-frame shifts in series of CCD-frames during the observation are commensurate with the errors in measuring their positions. These objects have velocities of apparent motion between CCD-frames not exceeding three RMS errors ($3\sigma$) of measurements of their positions. About 15\% of objects have a near-zero apparent motion in CCD-frames, including the objects beyond the Jupiter's orbit as well as the asteroids heading straight to the Earth. \par
\setlength{\parindent}{4ex}
The proposed method for detection of the object's near-zero apparent motion in series of CCD-frames is based on the Fisher f-criterion instead of using the traditional decision rules that are based on the maximum likelihood criterion. We analyzed the quality indicators of detection of the object's near-zero apparent motion applying statistical and in situ modeling techniques in terms of the conditional probability of the true detection of objects with a near-zero apparent motion. \par
\setlength{\parindent}{4ex}
The efficiency of method being implemented as a plugin for the Collection Light Technology (CoLiTec) software for automated asteroids and comets detection has been demonstrated. Among the objects discovered with this plugin, there was the sungrazing comet C/2012 S1 (ISON). Within 26 minutes of the observation, the comet's image has been moved by three pixels in a series of four CCD-frames (the velocity of its apparent motion at the moment of discovery was equal to 0.8 pixels per CCD-frame; the image size on the frame was about five pixels). Next verification in observations of asteroids with a near-zero apparent motion conducted with small telescopes has confirmed an efficiency of the method even in bad conditions (strong backlight from the full Moon). So, we recommend applying the proposed method for series of observations with four or more frames.}

\keywords{Methods: analytical, data analysis, numerical, statistical; Techniques: image processing; Comets: general, individual: ISON; Minor planets, asteroids: general, individual: 166, 1917, 1980, 3288, 6063, 20460, 138846, 242211}
\maketitle

\section{Introduction}
Different types of objects are detected in series of CCD-frames during observations: solar system minor bodies (SSOs); stars and large-scale diffuse sources (non-SSOs); charge transfer tails from bright stars, bright streaks from satellites, and noise sources amongst others. The difference between the detected SSOs and non-SSOs is that the non-SSOs have a zero velocity apparent motion on a set of frames, while the SSOs have a non-zero one. Wherein, a rapid detection of the objects with a near-zero velocity apparent motion both from the main belt of asteroids and beyond the Jupiter's orbit is very important for the asteroid-comet hazard problem as well as for the earliest recording new SSOs.

Over the past few decades, several powerful software tools and methods had been developed, allowing discovery and cataloging of thousands of SSOs (asteroids, comets, trans-Neptunians, Centaurs, etc.).
First of all, it was the Lincoln Near-Earth Asteroid Research (LINEAR) project \citep[][]{stokes}, which outperformed all asteroid search programs acted until 1998. This project brought the number of discovered SSOs to over 230~000, including 2~423 near-Earth objects (NEOs) and 279 comets \citep[][]{stokes1}. 
The second biggest asteroid survey, the Catalina Sky Survey \citep[][]{CSS}, started in 2005 as a search program for any potentially hazardous minor planets and allowed to discover more than 6~500 NEOs. The same program in the southern hemisphere, the Siding Spring Survey (SSS), was closed in 2013. 

A successful operation of these programs has stimulated new instruments and advanced CCD-cameras manufacturing as well as the development of new methods and algorithms for image processing and detecting faint SSOs. These methods of the automated search for very faint objects in a CCD-frame series were based, mostly, on the matched filter or the combined multiple frames along the typical SSO's motion \citep[][]{yanagisawa}. For example, the implementation of a multi-hypothesis velocity matched filter for LINEAR archive of images has produced about 25 \% new detections (mostly of faint SSOs) that were missed at the stage of a primary processing of observations \citep[][]{shucker}. Another algorithm, the interacting multiple model (IMM), was introduced as a modification of matched filter and provided a new structure for effective management of multiple filter models, while the selected parameters must be considered for the IMM optimizing \citep[][]{genovese}. 

The Panoramic Survey Telescope and Rapid Response System (Pan-STARRS) for surveying the sky for moving objects on a continual basis was designed as an array of four telescopes. The first telescope, PS1, is in a full operation since 2010 and is able to observe objects down to 22.5${^m}$ apparent magnitude. With the help of PS1 more than 2~860 NEOs and many comets have already been discovered \citep[see][]{henry}. PS1 uses the Moving Object Pipeline System, MOPS \citep{heasley}, which includes some methods and techniques for searching for the extremely faint and distant Sedna-like objects \citep{jedicke}, such as for example the modified intra-nightly linking algorithm, which includes a partial Hough transform method for quickly identifying of the multiple detections and post-processing step for intra-nightly linking (see \citet{parker}, \citet{myers}). 

These methods were successfully tested for simulations of processing the moving objects with MOPS on the Pan-STARRS and the Large Synoptic Survey Telescope, LSST \citep[][]{barnard}, the latter will be provided by the same pipeline system as on the Pan-STARRS \citep[][]{myers}. It is important that PS1 is a highly effective for discovering objects that could actually impact the Earth next 100 years \citep[][]{jedicke} and was complemented with the infrared data of the former WISE orbital telescope \citep[][]{dailey}.

In 2009, the authors of this paper developed the CoLiTec (Collection Light Technology) software for the automated detection of the solar system minor bodies in CCD-frames series (see, in detail, \url{http://www.neoastrosoft.com}
and papers by \citet{savanevych2,savanevych3,savanevych4,savanevych,vavilova1,vavilova,vavilova2,vavilova3,pohorelov}. Since 2009 it has been installed at several observatories: Andrushivka Astronomical Observatory (A50, Ukraine) \citep[][]{ivashchenko}, ISON-NM Observatory (H15, the US) \citep[][]{elenin}, ISON-Kislovodsk Observatory (D00, Russia) \citep[][]{kislovodsk}, ISON-Ussuriysk Observatory (C15, Russia) \citep[][]{elenin1}, Odessa-Mayaki (583, Ukraine) \citep[][]{troianskyi}, Vihorlat Observatory (Slovakia) \citep[][]{dubovsky}. 

The preliminary object's detection with CoLiTec software is based on the accumulation of the energy of signals along possible object tracks in a series of CCD-frames. Such accumulation is reached by the method of the multivalued transformation of the object coordinates that is equivalent to the Hough transformation \citep{savanevych3,savanevych4}. In general, CoLiTec software allows detecting of the objects with different velocities of the apparent motion by individual plugins for fast and slow objects, and objects with the near-zero apparent motion. CoLiTec software is widely used in a number of observatories. In total, four comets (C/2011 X1 (Elenin), P/2011 NO1 (Elenin), C/2012 S1 (ISON) and P/2013 V3 (Nevski)) and more than 1560 asteroids including 5 NEOs, 21 Trojan Jupiter asteroids and one Centaur were discovered using CoLiTec software as well as more than 700 000 positional CCD-measurements were sent to the Minor Planet Center \citep{ivashchenko,elenin,elenin1,savanevych}. Our comparison of statistical characteristics of positional CCD-measurements with CoLiTec and Astrometrica \citep[http://www.astrometrica.at;][]{miller,raab} software in the same set of test CCD-frames has demonstrated that the limits for reliable positional CCD-measurements with CoLiTec software are wider than those with Astrometrica one, in particular, for the area of extremely low signal-to-noise ratio (S/N) \citep[][]{savanevych}.

Besides the requirement of large computational effort \citep[see][]{shucker}, the main disadvantage of all the above mentioned methods implemented into software is a neglecting of near-zero apparent motion of objects in CCD-frames that has yet to be described and tested. So, the aim of this paper is to introduce a new computational method for detection of SSOs with a near-zero velocity of apparent motion in a series of CCD-frames. We propose considering these SSOs as a separate subclass, which includes objects whose inter-frame shifts during the observational session are commensurate with the errors in measuring their positions. We call the maximum permissible velocity of a near-zero apparent motion as $\varepsilonup$-velocity. Then, a subclass of SSOs with a near-zero apparent motion includes such SSOs, which have velocities of apparent motion between CCD-frames that are not exceed three RMS errors, $3\sigma$, of measurements of their positions ($\varepsilonup=3\sigma$). We will also use the notation of $3\sigma$-velocity instead of $\varepsilonup$-velocity to describe a near-zero apparent motion of SSOs.

The economy in the observational search resource leads to a reduction in the time between CCD-frames. This, in turn, leads to the fact that a significant part of SSOs will have an $\varepsilonup$-velocity apparent motion, in other words, have a shift, which is commensurate with the errors in estimating of their position. In general, there are about 15\% of SSOs with $\varepsilonup$-velocity motion. They are the objects beyond the Jupiter's orbit as well as asteroids moving to the observer along the view axis (heading straight to the Earth). Of course, when such an object is close enough, a parallax from the Earth's rotation will introduce a significant transverse motion that can be detectable. The proposed method allows us to locate objects with a near-zero apparent motion, including the potentially dangerous objects, at larger distances from the Earth than trivial methods. It gives more time to study such objects and to warn about their approach to the Earth in case of their hazardous behavior.

The structure of our paper is as follows. We describe a problem statement, a model of the apparent motion and hypothesis verification in Chapter 2. The task solution and new method are described in Chapter 3. Analysis of quality indicators of near-zero motion detection is provided in Chapter 4. Concluding remarks and discussion are given in Chapter 5. A mathematical rationale of the method is described in Appendices A-C.

\section{Problem statement}
The apparent motion of any object may be represented as the projection of its trajectory on the focal plane of a telescope. It is described by the model of rectilinear and uniform motion of an object along each coordinate independently during the tracking and formation of the series of its CCD-measurements (see Appendix A). 

Objects with significant apparent motion are easily detected by any methods of the trajectory determination, for example, the methods for inter-frame processing \citep{garcia,gong,vavilova}. The problem arises when we would like to detect an object with a near-zero apparent motion in CCD-frame series. Such an object can be falsely identified as the object with a $3\sigma$-velocity. 

The first step for solving this problem is a formation of the set of measurements $\Omega_{set}$ (A.5) (no more than one measurement per frame) for the object, which was preliminarily assigned to the objects with $3\sigma$-velocities. In its turn, such objects should be registered in the internal catalog of objects that are motionless in the series of CCD-frames \citep[][]{vavilova1}. This catalog is also helpful to reduce the number of false SSO detections in the software for automatic CCD-frame processing of asteroid surveys \citep[][]{pohorelov}.

In other words, the hypothesis $H_{0}$ that a certain set $\Omega_{set}$ (A.5) of measurements complies to the objects with a $3\sigma$-velocity is as follows: 
\begin{equation}
        \label{eq6}
        H_{0}:\sqrt{V_{x}^2+V_{y}^2}=0,
\end{equation}
where $V_{x}$, $V_{y}$ are the apparent velocities of object along each coordinate.

Then the more complicative alternative $H_{1}$ that the object with the set of measurements $\Omega_{set}$ (A.5) has a $3\sigma$-velocity will be written as:
\begin{equation}
        \label{eq7}
        H_{1}:\sqrt{V_{x}^2+V_{y}^2}>0.
\end{equation}
The false detection of the near-zero apparent motion of the object is an error of the first kind $\alpha$ assuming the validity of $H_{0}$ hypothesis (1). The skipping of the object with a $3\sigma$-velocity is an error of the second kind $\betaup$ under condition that the alternative hypothesis $H_{1}$ (2) is true. It is accepted in the community that the conditional probabilities of errors of the first $\alpha$ kind (conditional probability of the false detection, CPFD) and the second $\betaup$ kind (skipping of the object) are the indicators of a good quality detection \citep[][]{kuzmyn}. We also used the conditional probability of the true detection (CPTD) as a complement to the conditional probability of an error of the second $\betaup$ kind to unity ($1-\betaup$). 

So, the task solution may be formulated as follows: 1) it is necessary to develop computational methods for detecting the near-zero apparent motion of the object based on the analysis of a set $\Omega_{set}$ of measurements (A.5) obtained from a series of CCD-frames; 2) computational methods have to check the competing hypotheses of zero $H_{0}$ (1) and near-zero $H_{1}$ (2) apparent motion of the object.

\textbf{Maximum likelihood criterion.} Usually, hypotheses such as $H_{0}$ (6) and $H_{1}$ (7) are tested according to a maximum likelihood criterion \citep[][]{masson}\citep[][]{myung}, \citep[][]{miura}, \citep[][]{sandersreed} or any other criterion of the Bayesian group \citep[][]{lee}. The sufficient statistic for all the criteria is the likelihood ratio (LR), which is compared with critical values that are selected according to the specific criteria \citep[][]{morey}. If there are no opportunities to justify the a priori probabilities of hypotheses and losses related to wrong decisions, the developer can use either a maximum likelihood criteria or Neyman-Pearson approach \citep[][]{lee}. The unknown parameters of the likelihood function are evaluated by the same sample in which the hypotheses are tested. In mathematical statistics, such rules are called "substitutional rules for hypothesis testing" \citep[][]{lehman,morey}. In the technical literature, such rules are called "detection-measurement" \citep[][]{morey}.

The "detection" procedure precedes the "measurement" procedure for the substitutional decision rule. And this is a general principle for solving the problem of mixed optimization with discrete and continuous parameters \citep[][]{arora}. The decision statistics of hypotheses that correspond to different values of discrete parameters are compared with each other after the optimization of conditional likelihood functions for the value of their continuous parameters.
The software developers use the substitution rule of maximum likelihood despite the fact that the evidence is not proved mathematically. It should be compared with any new methods of hypothesis testing with a priori parametric uncertainty \citep[][]{gunawan}. The quality indicators of hypothesis testing can be examined only by statistical modeling or on the training samples of large experimental datasets. 

A likelihood function for detection of a near-zero apparent motion can be defined as the common density distribution of measurements of the object positions in a set of measurements (see Appendix B). Ordinary least square (OLS) evaluation of the parameters of the object's apparent motion as well as the variance of the object's positional estimates in a set of measurements are described in Appendix C. Using these parameters, one can obtain the maximum allowable (critical) value of the LR estimate for the detection of a near-zero apparent motion for the substitutional methods (C.11 - C.13).

\section{Task solution}
\textbf{Conversion of testing the hypothesis $H_{1}$ to the problem of validation of the statistical significance factor of the apparent motion.} One of the disadvantages of substitutional methods based on maximum likelihood criteria \citep{masson,myung} is the insufficient justification of their application when some parameters of likelihood function are unknown. The second one leads to the necessity of selecting the value of boundary decisive statistics \citep{miura,sandersreed}. Moreover, in our case, the substitutional methods are inefficient when the object's apparent motion is near-zero. 

Models (A.1) and (A.2) of the independent apparent motion along each coordinate are the classical models of linear regression with two parameters (start position and the velocity along each coordinate). Thus, in our case, the alternative $H_{1}$ hypothesis (2) about the object to be the SSO with a near-zero apparent motion is identical to the hypothesis about the statistical significance of the apparent motion.
We propose to check the statistical significance of the entire velocity for detection of a $3\sigma$-velocity, which is equivalent to check the hypothesis $H_{1}$.

\textbf{A method for detection of the near-zero apparent motion using Fisher f-criterion.} We propose to check the statistical significance of the entire velocity of the apparent motion of the object using f-criterion. F-test should be applied, when variances of the positions in a set of measurements are unknown. It is based on the fact that the f-distribution does not depend on the distribution of positional errors in a set of measurements \citep[][]{phillips,johnson}. Furthermore, there are also tabulated values of the Fisher distribution statistics \citep[][]{burden,melard}.

The f-criterion to check the statistical significance of the entire velocity of the apparent motion is represented as \citep[][]{phillips}:
\begin{equation}
        \label{eq23}
        f(\Omega_{set}) = \frac{R_{0}^2 - R_{1}^2}{R_{1}^2} \frac{N_{mea} - r}{w},
\end{equation}
where $w=1$ is the number of factors of the linear regression model that are verified by the hypothesis. In our case, the factor is the velocity of the apparent motion;

$r$ is a rank of the plan matrix $F_{x}$ \citep{burden} ($rang(F_{x} = r \leq min(m, N_{mea}))$);
\begin{equation}
        \label{eq24}
        F_{x}=\left\|
        \begin{array}{cc}
                1 & \Delta_{\tau 1}=(\tau_{1}-\tau_{0}) \\
                ... & ...\\
                1 & \Delta_{\tau k}=(\tau_{k}-\tau_{0})\\
                ... & ...\\
                1 & \Delta_{\tau Nmea}=(\tau_{Nmea}-\tau_{0})
        \end{array}
        \right\|.
\end{equation}

The rank of the $F_{x}$ matrix defined by (4) is equal to two for the linear model of the motion along one coordinate because a number $m$ of the estimated parameters of the motion is equal to two. As the apparent motion occurs along two coordinates, the number $m$ of its estimated parameters is equal to four. Accordingly, the rank $r$ of the $F_{x}$ matrix is four because $r = m$.

The statistic (3) has a Fisher probability distribution with ($w$, $N_{mea} - r$) degrees of freedom \citep[][]{phillips}. Its distribution corresponds to the distribution of the ratio of two independent random variables with a chi-square distribution \citep[][]{park}, degrees of freedom $w,$ and $N_{mea} - r$. For example, let the number $N_{fr}$ of CCD-frames in a series of frames to be $N_{fr} = 4$, and each frame contains the measurement of the object's position. Hence, for two coordinates the number of measurements is $2N_{mea} = 8$, $w = 1$, and the rank $r$ of the matrix $F_{x}$ (4) is $r = 4$. Therefore, statistic (3) has a Fisher probability distribution with ($1$, $4$) degrees of freedom.

To determine the maximum allowable (critical) tabulated value of the Fisher distribution statistics, we have to use the predefined significance level $\alpha$. Its value is the conditional probability of the false detection, CPFD, of the near-zero apparent motion. For example, if $\alpha=10^{-3}$, the maximum allowable $f_{cr}$ value of the Fisher distribution statistics with ($1$, $4$) degrees of freedom is $f_{cr} = 74.13$ \citep[][]{melard}. 

After transformation, the method for detection of the near-zero apparent motion using Fisher f-criterion is represented as:
\begin{equation}
        \label{eq25}
        \frac{R_{0}^2 - R_{1}^2}{R_{1}^2} \geq \frac{w f_{cr}}{N_{mea} - r}.
\end{equation}

\section{Indicators of quality of the near-zero apparent motion detection}
\textbf{Number of experiments for statistical modeling.} Errors in statistical modeling are defined by estimates of conditional probabilities of the false detection $\gammaup_{0}$ (validity of the $H_{0}$ hypothesis ) and true detection $\gammaup_{1}$ (validity of the alternative $H_{1}$ using the critical values of the decision statistics after modeling the $H_{0}$ hypothesis).

In our research we assumed that the reasonable values of errors of experimental frequencies are equal to $\gammaup_{0accept}=\alpha/10$, $\gammaup_{1accept}=10^{-3}$. Their dependence on the number of experiments for the statistical modeling (under the condition of a validity of the hypothesis $H_{0}$ and the alternative $H_{1}$) is determined by the empirical formulas:
\begin{equation}
        \label{eq26}
        N_{0exp} = 10^2 / \gammaup_{0accept};
\end{equation}
\begin{equation}
        \label{eq27}
        N_{1exp} = 10^2 / \gammaup_{1accept} = 10^{-6}.
\end{equation}

\textbf{Preconditions and constants for the methods of the statistical and in situ modeling.} To study the indicators of quality of the near-zero apparent motion detection using substitutional methods (see, Appendix C and formulas C.11 - C.13) in maximum likelihood approach, the appropriate maximum allowable values $\lambda_{cr}$ should be applied. These values are determined in accordance with the predefined level of significance $\alpha$ in the modeling of the hypothesis $H_{0}$ ($V = 0$).

For the statistical and in situ modeling, where the method (5) was used, we applied the tabulated value $f_{cr}$ of the Fisher distribution statistics with ($w$, $N_{mea} - r$) degrees of freedom \citep[][]{phillips}. As an alternative, the critical value $f_{cr}$ is determined according to the predefined level of significance $\alpha$ in the modeling of the hypothesis $H_{0}$ ($V = 0$).
 Normally distributed random variables were modeled using the Ziggurat method \citep[][]{marsaglia}. All the methods for detection of the near-zero apparent motion were analyzed on the same data set.

The following values of constants were used: the significance level is taken as $\alpha = 10^{-3}$ and $\alpha = 10^{-4}$; the number $N_{fr}$ of frames in a series is equal to $N_{fr} = (4, 6, 8, 10, 15)$. For modeling $H_{1}$ ($V > 0$) hypothesis the velocity module $V$ of the apparent motion was defined in relative terms, namely, RMS error of measurement deviations of the object's position ($V = k\sigma$).

Here the coefficient is equal to $k = (0, 0.5, 1, 1.25, 1.5, 1.75, 2, 3, 4, 5, 10)$. Mathematical expectation of external estimation of positional RMS error is $m(\hat{\sigma}_{out}) = 0$ and its RMS error is $\sigma(\hat{\sigma}_{out}) = (0.15, 0.25)$. If $\alpha = 10^{-3}$, the maximum allowable tabulated value of the Fisher distribution statistics with ($1$, $4$) degrees of freedom is equal to $f_{cr} = 74.13$ and if $\alpha = 10^{-4}$, it is $f_{cr} = 241.62$ \citep[][]{melard}.

\textbf{A method of statistical modeling for analysis of indicators of quality of the near-zero apparent motion detection in a series of CCD-frames.} Conditional probability of the true detection (CPTD) is calculated in terms of the frequency of LR estimates $\hat{\lambda}(\Omega_{set})$, or $f(\Omega_{set})$ exceeding the maximum allowable values $\lambda_{cr}$, or $f_{cr}$ for all methods of near-zero apparent motion detection:
\begin{equation}
        \label{eq28}
        D_{true} = N_{exc} / N_{1exp},
\end{equation}
where $N_{exc}$ is the number of exceedings of the critical value $\lambda_{cr}$ for the substitutional methods of maximum likelihood or $f_{cr}$ for the method with f-criterion. CPTD estimation is determined for the various number of frames $N_{fr}$ and various values of the apparent motion velocity module $V$.

Figure 1 ($\alpha = 10^{-3}$) shows the curves of near-zero apparent motion detected by different methods: the Fisher f-criterion (5) method (curve 1); substitutional method for maximum likelihood detection using the known variance of the position measurements (C.12) (curve 2); and substitutional method for maximum likelihood detection using external estimation of RMS error (C.13) $\hat{\sigma}_{out} = 0.15$ (curve 3) and $\hat{\sigma}_{out} = 0.25$ (curve 4).
\begin{figure}
        \label{pic1_1}
        \centering
        \begin{tikzpicture}[scale=0.8]
                \label{pic1a}
                \small
                \draw (0,0) -- coordinate (x axis mid) (8,0) node[right] {$V$};
                \draw (0,0) -- coordinate (y axis mid) (0,5) node[above=0.1cm] {$D_{true}$};
                \foreach \pos/\label in {0,1/$0.5\sigma$,2/$1\sigma$,3/$1.25\sigma$,4/$1.5\sigma$,5/$1.75\sigma$,
                                                                                        6/$2\sigma$,7/$3\sigma$,8/$4\sigma$}
                        \draw (\pos,0pt) -- (\pos,-3pt) node [anchor=north]  {\label};
                \foreach \pos/\label in {0,1/0.2,2/0.4,3/0.6,4/0.8,5/1}
                        \draw (0pt,\pos) -- (-3pt,\pos) node [anchor=east]  {\label};
                \draw[thick, mark=*] plot [smooth,tension=0.4] coordinates{(0,0.012) (1,0.03) (2,0.1) (3,0.17) (4,0.28) (5,0.38) (6,0.63) (7,1.77) (8,3.17)};
                \draw[<-,thick] (6.8,1.5) -- (7.3,1) node[anchor=west] {1};
                \draw[thick,dash pattern=on 1pt off 3pt on 3pt off 3pt, mark=*] plot [smooth,tension=0.4] coordinates{(0,0.005) (1,0.05) (2,0.52) (3,1.18) (4,2.16) (5,3.23) (6,4.1) (7,4.89) (8,5)};
                \draw[<-,thick] (4.5,2.8) -- (3.8,3) node[anchor=east] {2}; 
                \draw[thick,dashed, mark=*] plot [smooth,tension=0.4] coordinates{(0,0.005) (1,0.042) (2,0.43) (3,0.98) (4,1.86) (5,2.88) (6,3.82) (7,4.89) (8,5)};
                \draw[<-,thick] (4.4,2.2) -- (4.9,1.8) node[anchor=west] {3};
                \draw[thick,dotted, mark=*] plot [smooth,tension=0.4] coordinates{(0,0.005) (1,0.013) (2,0.059) (3,0.11) (4,0.21) (5,0.32) (6,0.61) (7,2.37) (8,4.25)};
                \draw[<-,thick] (7.35,3) -- (6.7,3.1) node[anchor=east] {4}; 
        \end{tikzpicture}
        \begin{tabular}{C{5cm}} a) $N_{fr}=4$ \end{tabular} 
        \begin{tikzpicture}[scale=0.8]
                \label{pic1b}
                \small
                \draw (0,0) -- coordinate (x axis mid) (8,0) node[right] {$V$};
                \draw (0,0) -- coordinate (y axis mid) (0,5) node[above=0.1cm] {$D_{true}$};
                \foreach \pos/\label in {0,1/$0.5\sigma$,2/$1\sigma$,3/$1.25\sigma$,4/$1.5\sigma$,5/$1.75\sigma$,
                                                                                        6/$2\sigma$,7/$3\sigma$,8/$4\sigma$}
                        \draw (\pos,0pt) -- (\pos,-3pt) node [anchor=north]  {\label};
                \foreach \pos/\label in {0,1/0.2,2/0.4,3/0.6,4/0.8,5/1}
                        \draw (0pt,\pos) -- (-3pt,\pos) node [anchor=east]  {\label};
                \draw[thick, mark=*] plot [smooth,tension=0.4] coordinates{(0,0.015) (1,0.24) (2,1.79) (3,3.07) (4,4.1) (5,4.7) (6,4.9) (7,5) (8,5)};
                \draw[<-,thick] (3.3,3.4) -- (3.9,3) node[anchor=west] {1};
                \draw[thick,dash pattern=on 1pt off 3pt on 3pt off 3pt, mark=*] plot [smooth,tension=0.4] coordinates{(0,0.005) (1,0.39) (2,3.67) (3,4.75) (4,4.98) (5,4.99) (6,5) (7,5) (8,5)};
                \draw[<-,thick] (2.4,4.3) -- (1.8,4.5) node[anchor=east] {2}; 
                \draw[thick,dashed, mark=*] plot [smooth,tension=0.4] coordinates{(0,0.005) (1,0.33) (2,3.36) (3,4.62) (4,4.96) (5,4.99) (6,5) (7,5) (8,5)};
                \draw[<-,thick] (2.4,4) -- (2.9,3.7) node[anchor=west] {3};
                \draw[thick,dotted, mark=*] plot [smooth,tension=0.4] coordinates{(0,0.005) (1,0.048) (2,0.47) (3,1.05) (4,1.91) (5,2.93) (6,3.83) (7,4.88) (8,5)}; 
                \draw[<-,thick] (4.3,2.2) -- (4.9,1.8) node[anchor=west] {4};
        \end{tikzpicture}
        \begin{tabular}{C{5cm}} b) $N_{fr}=6$ \end{tabular} 
        \begin{tikzpicture}[scale=0.8]
                \label{pic1c}
                \small
                \draw (0,0) -- coordinate (x axis mid) (8,0) node[right] {$V$};
                \draw (0,0) -- coordinate (y axis mid) (0,5) node[above=0.1cm] {$D_{true}$};
                \foreach \pos/\label in {0,1/$0.5\sigma$,2/$1\sigma$,3/$1.25\sigma$,4/$1.5\sigma$,5/$1.75\sigma$,
                                                                                        6/$2\sigma$,7/$3\sigma$,8/$4\sigma$}
                        \draw (\pos,0pt) -- (\pos,-3pt) node [anchor=north]  {\label};
                \foreach \pos/\label in {0,1/0.2,2/0.4,3/0.6,4/0.8,5/1}
                        \draw (0pt,\pos) -- (-3pt,\pos) node [anchor=east]  {\label};
                \draw[thick, mark=*] plot [smooth,tension=0.4] coordinates{(0,0.019) (1,3.58) (2,4.8) (3,5) (4,5) (5,5) (6,5) (7,5) (8,5)};
                \draw[<-,thick] (1.5,4.3) -- (1.4,3.6) node[anchor=north] {1};
                \draw[thick,dash pattern=on 1pt off 3pt on 3pt off 3pt, mark=*] plot [smooth,tension=0.4] coordinates{(0,0.005) (1,4.19) (2,4.9) (3,5) (4,5) (5,5) (6,5) (7,5) (8,5)};
                \draw[<-,thick] (1.25,4.3) -- (0.6,4.4) node[anchor=east] {2}; 
                \draw[thick,dashed, mark=*] plot [smooth,tension=0.4] coordinates{(0,0.005) (1,3.9) (2,4.9) (3,5) (4,5) (5,5) (6,5) (7,5) (8,5)};
                \draw[<-,thick] (1.4,4.65) -- (0.8,4.9) node[anchor=east] {3}; 
                \draw[thick,dotted, mark=*] plot [smooth,tension=0.4] coordinates{(0,0.005) (1,0.63) (2,4.31) (3,4.93) (4,4.99) (5,5) (6,5) (7,5) (8,5)};
                \draw[<-,thick] (1.4,2) -- (2,1.5) node[anchor=west] {4}; 
        \end{tikzpicture}
        \begin{tabular}{C{5cm}} c) $N_{fr}=10$ \end{tabular} 
        \begin{tikzpicture}[scale=0.8]
                \label{pic1d}
                \small
                \draw (0,0) -- coordinate (x axis mid) (8,0) node[right] {$V$};
                \draw (0,0) -- coordinate (y axis mid) (0,5) node[above=0.1cm] {$D_{true}$};
                \foreach \pos/\label in {0,1/$0.5\sigma$,2/$1\sigma$,3/$1.25\sigma$,4/$1.5\sigma$,5/$1.75\sigma$,
                                                                                        6/$2\sigma$,7/$3\sigma$,8/$4\sigma$}
                        \draw (\pos,0pt) -- (\pos,-3pt) node [anchor=north]  {\label};
                \foreach \pos/\label in {0,1/0.2,2/0.4,3/0.6,4/0.8,5/1}
                        \draw (0pt,\pos) -- (-3pt,\pos) node [anchor=east]  {\label};
                \draw[thick, mark=*] plot [smooth,tension=0.4] coordinates{(0,0.05) (1,4.45) (2,5) (3,5) (4,5) (5,5) (6,5) (7,5) (8,5)};
                \draw[<-,thick] (0.6,3) -- (0.3,3.6) node[anchor=south] {1}; 
                \draw[thick,dash pattern=on 1pt off 3pt on 3pt off 3pt, mark=*] plot [smooth,tension=0.4] coordinates{(0,0.005) (1,4.45) (2,5) (3,5) (4,5) (5,5) (6,5) (7,5) (8,5)};
                \draw[<-,thick] (1.3,4.8) -- (0.6,4.7) node[anchor=east] {2}; 
                \draw[thick,dashed, mark=*] plot [smooth,tension=0.4] coordinates{(0,0.005) (1,4.25) (2,5) (3,5) (4,5) (5,5) (6,5) (7,5) (8,5)};
                \draw[<-,thick] (1.3,4.6) -- (1.8,4.5) node[anchor=west] {3}; 
                \draw[thick,dotted, mark=*] plot [smooth,tension=0.4] coordinates{(0,0.005) (1,3.9) (2,4.9) (3,5) (4,5) (5,5) (6,5) (7,5) (8,5)}; 
                \draw[<-,thick] (1.2,4.2) -- (1.8,3.8) node[anchor=west] {4}; 
        \end{tikzpicture}
        \begin{tabular}{C{5cm}} d) $N_{fr}=15$ \end{tabular} 
        \caption{Curves of the near-zero apparent motion detection obtained by the method using Fisher f-criterion (1), substitutional methods with the known variance (2), with external estimations of RMS error 0.15 (3) and RMS error 0.25 (4)}
\end{figure}
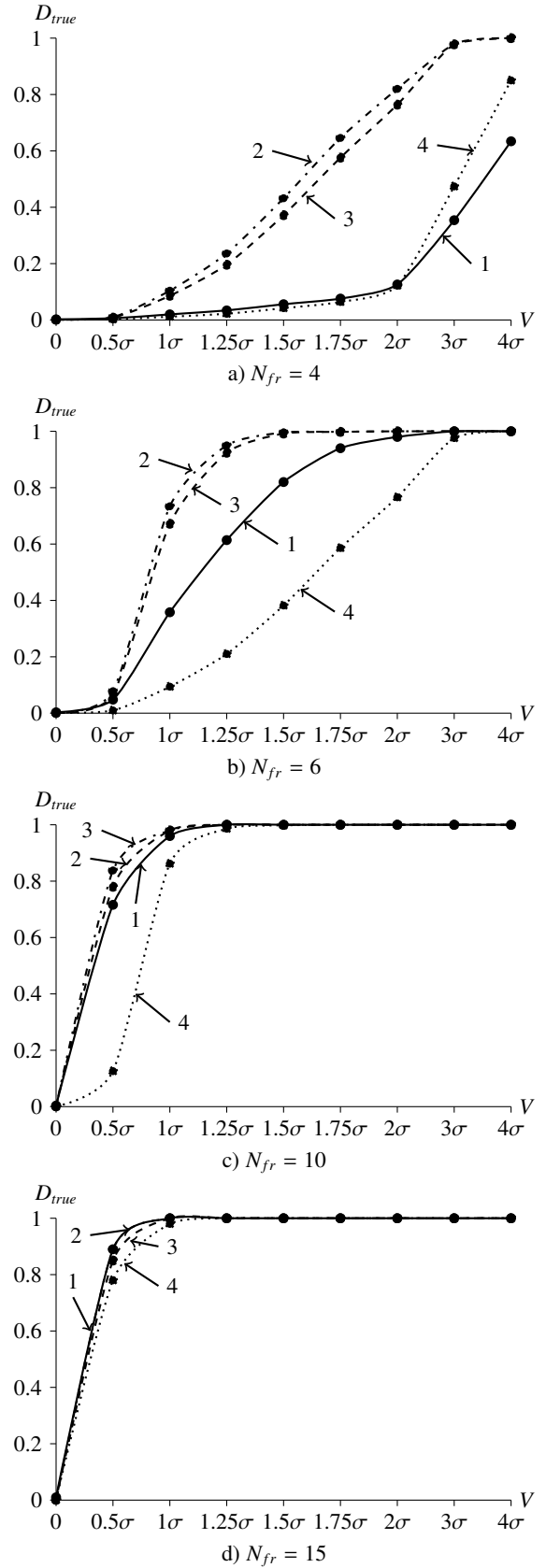

Figure 2 ($\alpha = 10^{-3}$) shows the curves of near-zero apparent motion detection obtained by the Fisher f-criterion method (5) with the critical tabulated value $f_{cr}$ of the Fisher distribution statistics with ($w$, $N_{mea} - r$) degrees of freedom \citep[][]{phillips} and the critical value $f_{cr}$ according to the predefined significance level $\alpha$.
\begin{figure}
        \label{pic1_3}
        \centering
        \begin{tikzpicture}[scale=0.8]
                \label{pic2a}
                \small
                \draw (0,0) -- coordinate (x axis mid) (8,0) node[right] {$V$};
                \draw (0,0) -- coordinate (y axis mid) (0,5) node[above=0.1cm] {$D_{true}$};
                \foreach \pos/\label in {0,1/$0.5\sigma$,2/$1\sigma$,3/$1.25\sigma$,4/$1.5\sigma$,5/$1.75\sigma$,
                                                                                        6/$2\sigma$,7/$3\sigma$,8/$4\sigma$}
                        \draw (\pos,0pt) -- (\pos,-3pt) node [anchor=north]  {\label};
                \foreach \pos/\label in {0,1/0.2,2/0.4,3/0.6,4/0.8,5/1}
                        \draw (0pt,\pos) -- (-3pt,\pos) node [anchor=east]  {\label};
                \draw[thick,dashed, mark=*] plot [smooth,tension=0.4] coordinates{(0,0.005) (1,0.012) (2,0.04) (3,0.07) (4,0.12) (5,0.19) (6,0.28) (7,0.92) (8,1.94)};
                \draw[thick, mark=*] plot [smooth,tension=0.4] coordinates{(0,0.012) (1,0.03) (2,0.1) (3,0.17) (4,0.28) (5,0.38) (6,0.63) (7,1.77) (8,3.17)};
        \end{tikzpicture}
        \begin{tabular}{C{5cm}} a) $N_{fr}=4$ \end{tabular} 
        \begin{tikzpicture}[scale=0.8]
                \label{pic2b}
                \small
                \draw (0,0) -- coordinate (x axis mid) (8,0) node[right] {$V$};
                \draw (0,0) -- coordinate (y axis mid) (0,5) node[above=0.1cm] {$D_{true}$};
                \foreach \pos/\label in {0,1/$0.5\sigma$,2/$1\sigma$,3/$1.25\sigma$,4/$1.5\sigma$,5/$1.75\sigma$,
                                                                                        6/$2\sigma$,7/$3\sigma$,8/$4\sigma$}
                        \draw (\pos,0pt) -- (\pos,-3pt) node [anchor=north]  {\label};
                \foreach \pos/\label in {0,1/0.2,2/0.4,3/0.6,4/0.8,5/1}
                        \draw (0pt,\pos) -- (-3pt,\pos) node [anchor=east]  {\label};
                \draw[thick,dashed, mark=*] plot [smooth,tension=0.4] coordinates{(0,0.005) (1,0.087) (2,0.93) (3,1.9) (4,3) (5,3.99) (6,4.59) (7,4.99) (8,5)};
                \draw[thick, mark=*] plot [smooth,tension=0.4] coordinates{(0,0.015) (1,0.24) (2,1.79) (3,3.07) (4,4.1) (5,4.7) (6,4.9) (7,5) (8,5)};
        \end{tikzpicture}
        \begin{tabular}{C{5cm}} b) $N_{fr}=6$ \end{tabular} 
        \begin{tikzpicture}[scale=0.8]
                \label{pic2c}
                \small
                \draw (0,0) -- coordinate (x axis mid) (8,0) node[right] {$V$};
                \draw (0,0) -- coordinate (y axis mid) (0,5) node[above=0.1cm] {$D_{true}$};
                \foreach \pos/\label in {0,1/$0.5\sigma$,2/$1\sigma$,3/$1.25\sigma$,4/$1.5\sigma$,5/$1.75\sigma$,
                                                                                        6/$2\sigma$,7/$3\sigma$,8/$4\sigma$}
                        \draw (\pos,0pt) -- (\pos,-3pt) node [anchor=north]  {\label};
                \foreach \pos/\label in {0,1/0.2,2/0.4,3/0.6,4/0.8,5/1}
                        \draw (0pt,\pos) -- (-3pt,\pos) node [anchor=east]  {\label};
                \draw[thick,dashed, mark=*] plot [smooth,tension=0.4] coordinates{(0,0.005) (1,2.57) (2,4.7) (3,5) (4,5) (5,5) (6,5) (7,5) (8,5)};
                \draw[thick, mark=*] plot [smooth,tension=0.4] coordinates{(0,0.019) (1,3.58) (2,4.8) (3,5) (4,5) (5,5) (6,5) (7,5) (8,5)};
        \end{tikzpicture}
        \begin{tabular}{C{5cm}} c) $N_{fr}=10$ \end{tabular} 
        \begin{tikzpicture}[scale=0.8]
                \label{pic2d}
                \small
                \draw (0,0) -- coordinate (x axis mid) (8,0) node[right] {$V$};
                \draw (0,0) -- coordinate (y axis mid) (0,5) node[above=0.1cm] {$D_{true}$};
                \foreach \pos/\label in {0,1/$0.5\sigma$,2/$1\sigma$,3/$1.25\sigma$,4/$1.5\sigma$,5/$1.75\sigma$,
                                                                                        6/$2\sigma$,7/$3\sigma$,8/$4\sigma$}
                        \draw (\pos,0pt) -- (\pos,-3pt) node [anchor=north]  {\label};
                \foreach \pos/\label in {0,1/0.2,2/0.4,3/0.6,4/0.8,5/1}
                        \draw (0pt,\pos) -- (-3pt,\pos) node [anchor=east]  {\label};
                \draw[thick,dashed, mark=*] plot [smooth,tension=0.4] coordinates{(0,0.005) (1,4) (2,4.9) (3,5) (4,5) (5,5) (6,5) (7,5) (8,5)};
                \draw[thick, mark=*] plot [smooth,tension=0.4] coordinates{(0,0.05) (1,4.45) (2,5) (3,5) (4,5) (5,5) (6,5) (7,5) (8,5)};
        \end{tikzpicture}
        \begin{tabular}{C{5cm}} d) $N_{fr}=15$ \end{tabular} 
        \caption{The curves of the near-zero apparent motion detection obtained by the Fisher f-criterion method with the critical tabulated value (solid line) and the critical value according to the predefined significance level $\alpha$ (dashed line)}
\end{figure}

\textbf{A method of in situ modeling for analysis of indicators of quality of the near-zero apparent motion detection on a series of CCD-frames.} In this case, it is impossible to restore the real law of the errors' distribution completely. The method of in situ modeling is, therefore, more appropriate \citep[][]{kuzmyn}. 

We compiled the set of objects with practically zero apparent motion in the framework of the CoLiTec project \citep{savanevych,savanevych1} and used it as the internal catalog (IC) of motionless objects in a series of frames \citep[][]{vavilova1}. 

It is important to note that the objects exactly from the internal catalog were selected as in situ data. Because the positions of objects from this catalog are fixed, so deviations of their estimated positions from their average value can be regarded as evaluations of their errors. These values can be used in the in situ modeling. 

Further, these deviations should be added to the determined values of the object's displacements according to their velocities of the apparent motion. Thereby, it is possible to use the real laws of the positional errors distribution in the study of their motion by the in situ modeling method.

\textbf{In situ data.} Series of CCD-frames from observatories ISON-NM (MPC code - "H15") \citep[][]{molotov} and ISON-Kislovodsk (MPC code - "D00") \citep[][]{kislovodsk} were selected as the in situ data. The ISON-NM observatory is equipped with a 40 cm telescope SANTEL-400AN with CCD-camera FLI ML09000-65 (3056 x 3056 pixels, the pixel size is 12 microns). Exposure time was 150 seconds. 

The ISON-Kislovodsk observatory is equipped with a 19.2 cm wide-field telescope GENON (VT-78) with CCD-camera FLI ML09000-65 (4008 x 2672 pixels,the pixel size is 9 microns). Exposure time was 180 seconds. Figures 3 and 4 show the curves of the near-zero apparent motion detection obtained by the Fisher f-criterion (5) and by the substitutional method of maximum likelihood with an external estimation of RMS error (C.13) for two sources of in situ data.
\begin{figure}[H]
        \centering
        \begin{tikzpicture}[scale=0.8]
                \label{pic5a}
                \small
                \draw (0,0) -- coordinate (x axis mid) (10,0) node[right] {$V$};
                \draw (0,0) -- coordinate (y axis mid) (0,5) node[above=0.1cm] {$D_{true}$};
                \foreach \pos/\label in {0,1/$0.5\sigma$,2/$1\sigma$,3/$1.25\sigma$,4/$1.5\sigma$,5/$1.75\sigma$,
                                                                                        6/$2\sigma$,7/$3\sigma$,8/$4\sigma$,9/$5\sigma$,10/$10\sigma$}
                        \draw (\pos,0pt) -- (\pos,-3pt) node [anchor=north]  {\label};
                \foreach \pos/\label in {0,1/0.2,2/0.4,3/0.6,4/0.8,5/1}
                        \draw (0pt,\pos) -- (-3pt,\pos) node [anchor=east]  {\label};
                \draw[thick,dashed, mark=*] plot [smooth,tension=0.4] coordinates{(0,0.005) (1,0.008) (2,0.01) (3,0.015) (4,0.019) (5,0.026) (6,0.038) (7,0.26) (8,2) (9,4.7) (10,5)};
                \draw[thick, mark=*] plot [smooth,tension=0.4] coordinates{(0,0.01) (1,0.6) (2,1.8) (3,2.2) (4,2.5) (5,2.8) (6,3) (7,3.6) (8,4) (9,4.3) (10,5)};
        \end{tikzpicture}
        \begin{tabular}{C{5cm}} a) $\alpha = 10^{-3}$ \end{tabular} 
        \begin{tikzpicture}[scale=0.8]
                \label{pic5b}
                \small
                \draw (0,0) -- coordinate (x axis mid) (10,0) node[right] {$V$};
                \draw (0,0) -- coordinate (y axis mid) (0,5) node[above=0.1cm] {$D_{true}$};
                \foreach \pos/\label in {0,1/$0.5\sigma$,2/$1\sigma$,3/$1.25\sigma$,4/$1.5\sigma$,5/$1.75\sigma$,
                                                                                        6/$2\sigma$,7/$3\sigma$,8/$4\sigma$,9/$5\sigma$,10/$10\sigma$}
                        \draw (\pos,0pt) -- (\pos,-3pt) node [anchor=north]  {\label};
                \foreach \pos/\label in {0,1/0.2,2/0.4,3/0.6,4/0.8,5/1}
                        \draw (0pt,\pos) -- (-3pt,\pos) node [anchor=east]  {\label};
                \draw[thick,dashed, mark=*] plot [smooth,tension=0.4] coordinates{(0,0.005) (1,0.008) (2,0.01) (3,0.015) (4,0.019) (5,0.023) (6,0.025) (7,0.03) (8,0.23) (9,1.25) (10,5)};
                \draw[thick, mark=*] plot [smooth,tension=0.4] coordinates{(0,0.01) (1,0.15) (2,0.7) (3,1) (4,1.4) (5,1.6) (6,1.9) (7,2.7) (8,3.1) (9,3.5) (10,4.7)};
        \end{tikzpicture}
        \begin{tabular}{C{5cm}} b) $\alpha = 10^{-4}$ \end{tabular} 
        \caption{Curves of the near-zero apparent motion detection with the SANTEL-400AN telescope obtained by the Fisher f-criterion method (solid line) and by the substitutional method with external estimation of RMS error 0.15 (dashed line)}
\end{figure}
\begin{figure}[H]
        \centering
        \begin{tikzpicture}[scale=0.8]
                \label{pic6a}
                \small
                \draw (0,0) -- coordinate (x axis mid) (10,0) node[right] {$V$};
                \draw (0,0) -- coordinate (y axis mid) (0,5) node[above=0.1cm] {$D_{true}$};
                \foreach \pos/\label in {0,1/$0.5\sigma$,2/$1\sigma$,3/$1.25\sigma$,4/$1.5\sigma$,5/$1.75\sigma$,
                                                                                        6/$2\sigma$,7/$3\sigma$,8/$4\sigma$,9/$5\sigma$,10/$10\sigma$}
                        \draw (\pos,0pt) -- (\pos,-3pt) node [anchor=north]  {\label};
                \foreach \pos/\label in {0,1/0.2,2/0.4,3/0.6,4/0.8,5/1}
                        \draw (0pt,\pos) -- (-3pt,\pos) node [anchor=east]  {\label};
                \draw[thick, dashed, mark=*] plot [smooth,tension=0.4] coordinates{(0,0.005) (1,0.01) (2,0.03) (3,0.07) (4,0.1) (5,0.2) (6,0.7) (7,4.2) (8,4.8) (9,4.9) (10,5)};
                \draw[thick, mark=*] plot [smooth,tension=0.4] coordinates{(0,0.01) (1,1.1) (2,1.8) (3,2.0) (4,2.3) (5,2.5) (6,2.7) (7,3.3) (8,3.8) (9,4.2) (10,5)};
        \end{tikzpicture}
        \begin{tabular}{C{5cm}} a) $\alpha = 10^{-3}$ \end{tabular} 
        \begin{tikzpicture}[scale=0.8]
                \label{pic6b}
                \small
                \draw (0,0) -- coordinate (x axis mid) (10,0) node[right] {$V$};
                \draw (0,0) -- coordinate (y axis mid) (0,5) node[above=0.1cm] {$D_{true}$};
                \foreach \pos/\label in {0,1/$0.5\sigma$,2/$1\sigma$,3/$1.25\sigma$,4/$1.5\sigma$,5/$1.75\sigma$,
                                                                                        6/$2\sigma$,7/$3\sigma$,8/$4\sigma$,9/$5\sigma$,10/$10\sigma$}
                        \draw (\pos,0pt) -- (\pos,-3pt) node [anchor=north]  {\label};
                \foreach \pos/\label in {0,1/0.2,2/0.4,3/0.6,4/0.8,5/1}
                        \draw (0pt,\pos) -- (-3pt,\pos) node [anchor=east]  {\label};
                \draw[thick,dashed, mark=*] plot [smooth,tension=0.4] coordinates{(0,0.001) (1,0.002) (2,0.008) (3,0.017) (4,0.03) (5,0.06) (6,0.31) (7,1.37) (8,4.5) (9,4.9) (10,5)};
                \draw[thick, mark=*] plot [smooth,tension=0.4] coordinates{(0,0.002) (1,0.74) (2,1.2) (3,1.4) (4,1.6) (5,1.7) (6,1.9) (7,2.4) (8,2.8) (9,3.6) (10,4.7)};
        \end{tikzpicture}
        \begin{tabular}{C{5cm}} b) $\alpha = 10^{-4}$ \end{tabular}
        \caption{Curves of the near-zero apparent motion detection with the GENON (VT-78) telescope obtained by the Fisher f-criterion method (solid line) and by the substitutional method with external estimation of RMS error 0.15 (dashed line)}
\end{figure}
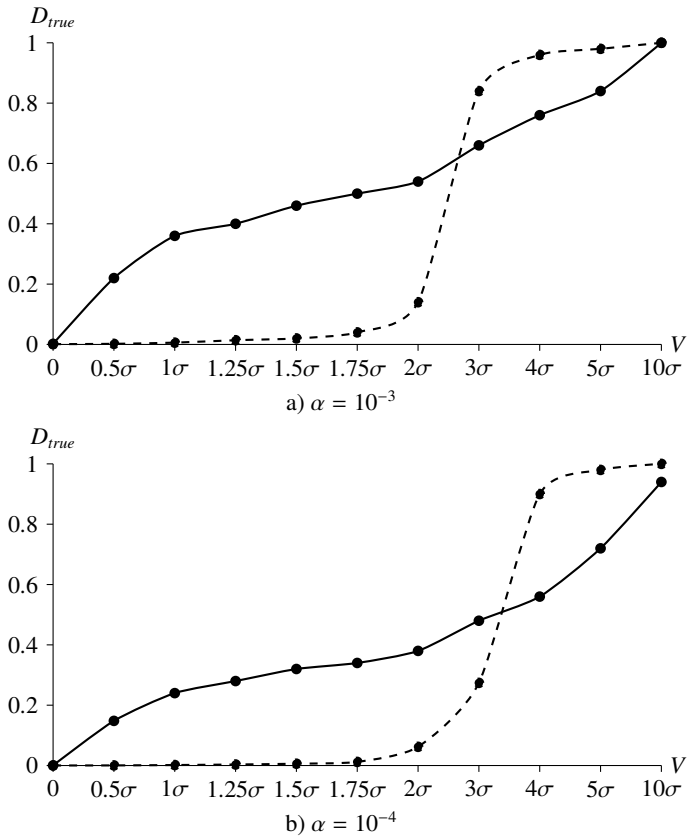

\textbf{Analysis of indicators of quality of the near-zero apparent motion detection in a series of CCD-frames by the method of statistical modeling.} Analyzing different approaches, we can note that the substitutional methods of maximum likelihood detection with known variance of the object's position (C.12) depicted by the curve 2 in Fig. 1, and the methods with external estimation of RMS errors $\hat{\sigma}_{out} = 0.15$ (C.13) represented by the curve 3 in the same figure are the most sensitive to the object velocity changes. For example, CPTD of the near-zero apparent motion for these methods increases in the series consisting of four frames and having $V = 0.5\sigma$. Here, $\sigma$ is an RMS error of the errors of estimated positions. For other methods the velocity module of the apparent motion is not less than $V = 1.25\sigma$, and if $N_{fr} = 6$, not less than $V = \sigma$.
 
The curve 1 in Fig. 1 demonstrates that the near-zero apparent motion detection method with Fisher f-criterion (5) is not effective enough with the data of statistical modeling, when the number of frames $N_{fr}$ is small. But if $N_{fr}$ is not less than eight, this method is not inferior to other ones by CPTD. In own turn, the substitutional method of maximum likelihood with the known variance of the object's position (C.12) exists only in theory and can not be applied in practice.

Hereby, the substitutional method of maximum likelihood with external estimation of RMS error (C.13) described by curve 3 in Fig. 1 is the most effective and flexible. We remember that the external estimation can be obtained from measurements of the other objects in CCD-frame.

On the other hand, the determination of critical values for all substitutional methods encounters formidable obstacles. First of all, it is not clear how to separate a set of stars (objects with a zero rate motion) from the objects with a near-zero apparent motion to determine them. Also, this process is very time- and resource-consuming and difficult to apply in rapidly changing conditions of observations in modern asteroid surveys.

In statistical modeling, the critical values $f_{cr}$ of the f-criterion determined according to the predefined significance levels are almost equal to the tabulated critical values of Fisher distribution statistics with ($w$, $N_{mea} - r$) degrees of freedom \citep[][]{phillips,melard} of the method (5). It is obviously seen in Fig. 2. Moreover, these figures demonstrate that the similarity of these critical values of decisive statistic does not depend on the number of frames in the series. 

Hence, it is not necessary to determine them for the different number of frames $N_{fr}$ and observation conditions. It is enough to use the maximum allowable tabulated value \citep[][]{melard}.

Following from our statistical experiments, we can note that the method for the near-zero apparent motion detection with Fisher f-criterion (5) is more effective for the large number of CCD-frames and the velocity module of the apparent motion $V = 0.5\sigma$ as it's seen in Fig. 2.

\textbf{Analysis of indicators of quality of the near-zero apparent motion detection in a series of CCD-frames by the method of in situ modeling.} It is found that the method for detection of the object's near-zero apparent motion using Fisher f-criterion (5) is the most sensitive to changes in the object's velocity (Fig. 3, 4). As shown earlier, CPTD for this method increases when series includes four frames or more and when $V = 0.5\sigma$. For other methods the velocity module of the apparent motion should be not less than $V = 1.25\sigma$.

In addition, the method of the near-zero apparent motion detection using Fisher f-criterion (5) is stable and does not depend on the kind of telescope (Fig. 5a). Therefore, there is no need to undertake additional steps for determining the critical value of the decisive statistic after the equipment replacement or observational conditions change. Other methods of the apparent motion detection encounter problems when determining the critical values as it is obvious from Fig. 5b.
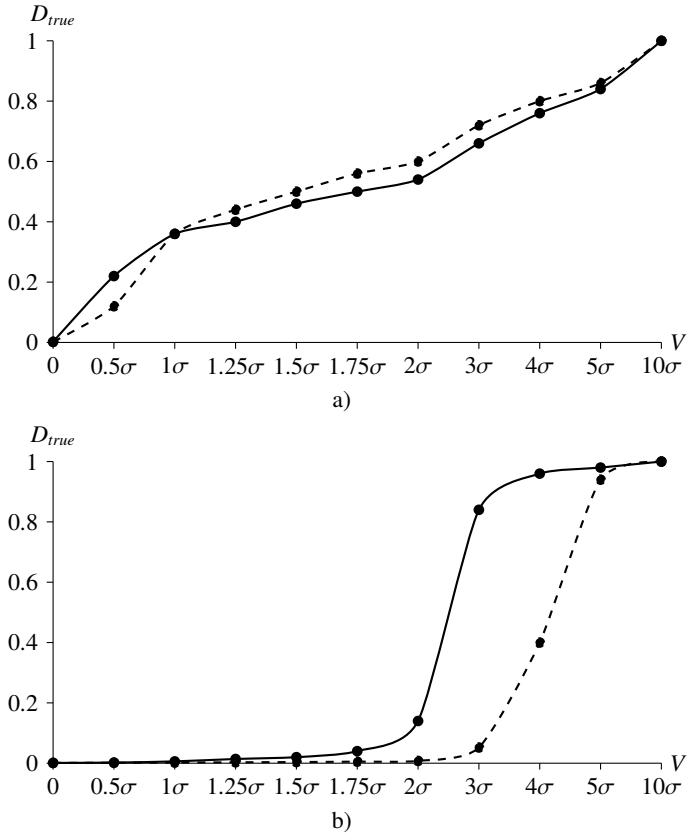
\begin{figure}
        \centering
        \begin{tikzpicture}[scale=0.8]
                \label{pic7a}
                \small
                \draw (0,0) -- coordinate (x axis mid) (10,0) node[right] {$V$};
                \draw (0,0) -- coordinate (y axis mid) (0,5) node[above=0.1cm] {$D_{true}$};
                \foreach \pos/\label in {0,1/$0.5\sigma$,2/$1\sigma$,3/$1.25\sigma$,4/$1.5\sigma$,5/$1.75\sigma$,
                                                                                        6/$2\sigma$,7/$3\sigma$,8/$4\sigma$,9/$5\sigma$,10/$10\sigma$}
                        \draw (\pos,0pt) -- (\pos,-3pt) node [anchor=north]  {\label};
                \foreach \pos/\label in {0,1/0.2,2/0.4,3/0.6,4/0.8,5/1}
                        \draw (0pt,\pos) -- (-3pt,\pos) node [anchor=east]  {\label};
                \draw[thick,dashed, mark=*] plot [smooth,tension=0.4] coordinates{(0,0.01) (1,0.6) (2,1.8) (3,2.2) (4,2.5) (5,2.8) (6,3) (7,3.6) (8,4) (9,4.3) (10,5)};
                \draw[thick, mark=*] plot [smooth,tension=0.4] coordinates{(0,0.01) (1,1.1) (2,1.8) (3,2.0) (4,2.3) (5,2.5) (6,2.7) (7,3.3) (8,3.8) (9,4.2) (10,5)};
        \end{tikzpicture}
        \begin{tabular}{C{5cm}} a) \end{tabular}
        \begin{tikzpicture}[scale=0.8]
                \label{pic7b}
                \small
                \draw (0,0) -- coordinate (x axis mid) (10,0) node[right] {$V$};
                \draw (0,0) -- coordinate (y axis mid) (0,5) node[above=0.1cm] {$D_{true}$};
                \foreach \pos/\label in {0,1/$0.5\sigma$,2/$1\sigma$,3/$1.25\sigma$,4/$1.5\sigma$,5/$1.75\sigma$,
                                                                                        6/$2\sigma$,7/$3\sigma$,8/$4\sigma$,9/$5\sigma$,10/$10\sigma$}
                        \draw (\pos,0pt) -- (\pos,-3pt) node [anchor=north]  {\label};
                \foreach \pos/\label in {0,1/0.2,2/0.4,3/0.6,4/0.8,5/1}
                        \draw (0pt,\pos) -- (-3pt,\pos) node [anchor=east]  {\label};
                \draw[thick,dashed, mark=*] plot [smooth,tension=0.4] coordinates{(0,0.005) (1,0.008) (2,0.01) (3,0.015) (4,0.019) (5,0.026) (6,0.038) (7,0.26) (8,2) (9,4.7) (10,5)};
                \draw[thick, mark=*] plot [smooth,tension=0.4] coordinates{(0,0.005) (1,0.01) (2,0.03) (3,0.07) (4,0.1) (5,0.2) (6,0.7) (7,4.2) (8,4.8) (9,4.9) (10,5)};
        \end{tikzpicture}
        \begin{tabular}{C{5cm}} b) \end{tabular}
        \caption{Curves of the near-zero apparent motion detection with the GENON (VT-78) (solid line) and SANTEL-400AN (dashed line) telescopes ($\alpha = 10^{-3}$) obtained by the Fisher f-criterion method (a), substitutional method for maximum likelihood detection with external estimation of RMS error (b)}
\end{figure}

\textbf{Examples of objects discovered by the method of near-zero apparent motion detection in a series of CCD-frames using significance criteria of the apparent motion.} There are many of objects with near-zero apparent motion that were detected by the CoLiTec software for automated asteroids and comets discoveries \citep[][]{savanevych1}. The plugin implements the method of detection using the Fisher f-criterion (5). Table 1 gives information about several observatories at which the CoLiTec software is installed.
\begin{table}[H]
        \label{table1}
        \caption{Information about observatories and telescopes at which the CoLiTec software is installed}
        \centering
        \scriptsize
        \begin{tabular}{| L{1.5cm} | C{1.3cm} | C{1.3cm} | C{1.4cm} | C{1.4cm} |}
                \hline
                Observatory & \multicolumn{2}{C{2.6cm} |}{ISON-Uzhgorod Observatory} & Cerro Tololo Inter-American Observatory (CTIO) & ISON-Kislovodsk Observatory \\ \hline
                MPC code & K99 & - & - & D00 \\ \hline
                Telescope & ChV-400 & BRC-250M & Promt8 & Santel-400AN \\ \hline
                Aperture, cm & 40 & 25 & 61 & 40 \\ \hline
                CCD-camera & FLI PL09000 & Apogee \newline Alta U9 & Apogee F42 & FLI ML09000-65 \\ \hline
                Resolution, pix & 3056 x 3056 & 3072 x 2048 & 2048 x 2048 & 3056 x 3056 \\ \hline
                Pixel size, $\mu$m & 12 & 9 & 13.5 & 12 \\ \hline
                Scale, " & 1.42 & 1.46 & 0.66 & 2.06 \\ \hline
        \end{tabular}
\end{table}
The real-life examples of detection of asteroids 1917, 6063, 242211, 3288 and 1980, 20460, 138846, 166 with a near-zero apparent motion are described in Tables 2 and 3 respectively. 

The observations were conducted in 2017 in the period from 3 to 19 July with different small telescopes and confirmed an efficiency of the method even in bad conditions (strong backlight from the full Moon).

Tables 2 and 3 contain the following apparent motion parameters of the aforementioned asteroids: 
date of observations;
name of telescope; 
exposure time during the observation; 
apparent velocities of object along each coordinate $\hat{V}_{x}$ and $\hat{V}_{y}$ in the rectangular coordinate system (CS) (see, Appendix C, formulas (C.1), (C.2); 
apparent velocities of objects $\hat{V}_{RA}$ and $\hat{V}_{DE}$ in the equatorial CS determined from the observational data; 
apparent velocities of object $\hat{V}_{RAcat}$ and $\hat{V}_{DEcat}$ in the equatorial CS determined from the Horizons system \citep[][]{giorgini} for the same times of observation; 
velocity module $\hat{V}$ of the apparent motion of object determined from the observational data ($\hat{V}=\sqrt{\hat{V}_{x}^2+\hat{V}_{y}^2}$); 
velocity module $\hat{V}_{cat}$ of the apparent motion of object determined from the Horizons system; 
average FWHM of object in five frames; 
average SNR of object in five frames; 
RMS error of stars positional estimates $\hat{\sigma}_{0}$ (C.7) from UCAC4 catalog \citep[][]{zacharias} with SNR approximately equal to the object's SNR; 
brightness ${Mag}_{cat}$ of the object determined from the Horizons system; 
angular distance between the observed asteroid and the Moon;
phase of the Moon, percentage illumination by the Sun;
coefficient of the velocity module $\hat{V}_{cat}$ of the apparent motion of object determined in relative terms, in other words, RMS error of measurement deviations of the object's position ($k = \hat{V}/\hat{\sigma}_{0}$).

\begin{table}[H]
        \label{table2a}
        \caption{Examples of asteroids 1917, 6063, 242211, 3288 with a near-zero apparent motion that were detected by the proposed method using Fisher f-criterion (5)}
        \centering
        \scriptsize
        \begin{tabular}{| L{2.2cm} | C{1.2cm} | C{1.2cm} | C{1.2cm} | C{1.2cm}  |}
                \hline
                Parameters $\diagdown$ Objects 
                & 1917            & 6063          & 242211        & 3288         \\ \hline
                Date of observation 
                & 2017-07-11      & 2017-07-11     & 2017-07-13    & 2017-07-19     \\ \hline
                Telescope 
                & Promt8         & Promt8        & Promt8        & Promt8         \\ \hline
                Exposure, s 
                & 80             & 40            & 40            & 20              \\ \hline
                $\hat{V}_{x}$, pix/fr 
                & 0.47           & 0.94          & -0.56         & 0.01           \\ \hline
                $\hat{V}_{y}$, pix/fr 
                & -0.47          & 0.73          & 0.36          & -0.47           \\ \hline
                $\hat{V}_{RA}$, "/fr 
                & -0.49          & 0.66          & -0.30         & -0.22           \\ \hline
                $\hat{V}_{DE}$, "/fr 
                & -0.25          & 0.65          & -0.39         & -0.02            \\ \hline
                $\hat{V}_{RAcat}$, "/fr 
                & -0.32          & 0.66          & -0.22         & -0.31           \\ \hline
                $\hat{V}_{DEcat}$, "/fr 
                & -0.34          & 0.65          & -0.37         & -0.04           \\ \hline
                $\hat{V}$, pix/fr 
                & \textbf{0.66}  & \textbf{1.19} & \textbf{0.67} & \textbf{0.50}    \\ \hline
                $\hat{V}$, "/fr 
                & 0.55           & 0.93          & 0.49          & 0.22             \\ \hline
                ${V}_{cat}$, "/fr 
                & 0.47           & 0.93          & 0.43          & 0.31              \\ \hline
                Average FWHM, pix 
                & 3.48            & 3.68          & 4.62          & 5.70             \\ \hline
                Average SNR, "/fr 
                & 6.86            & 10.04         & 12.83         & 11.86             \\ \hline
                $\hat{\sigma}_{0}$, pix (UCAC4) 
                & \textbf{0.40}   & \textbf{0.45} & \textbf{0.41} & \textbf{0.30}      \\ \hline
                $\hat{\sigma}_{0}$, " 
                & 0.30            & 0.19          & 0.28          & 0.20               \\ \hline
                ${Mag}_{cat}$, $^{m}$
                & 18.2            & 17.38         & 17.17         & 18.24              \\ \hline
                Asteroid-Moon dist., deg
                & 97              & 82.5          & 68            & 91.5               \\ \hline
                Moon phase \%
                & 91              & 91            & 76            & 14                  \\ \hline 
                $k=\hat{V} / \hat{\sigma}_{0}$ 
                & \textbf{1.65}   & \textbf{2.64} & \textbf{1.63} & \textbf{1.67}           \\ \hline
        \end{tabular}
\end{table}

\begin{table}[H]
\label{table2b}
\caption{Examples of asteroids 1980, 20460, 138846, 166 with a near-zero apparent motion that were detected by the proposed method using Fisher f-criterion (5)}
\centering
\scriptsize
\begin{tabular}{| L{2.2cm} | C{1.2cm} | C{1.2cm} | C{1.2cm} | C{1.2cm} |}
        \hline
        Parameters $\diagdown$ Objects 
        & 1980                          & 20460           & 138846        & 166\\ \hline
        Date of observation 
        & 2017-07-09                    & 2017-07-03       & 2017-07-13    & 2017-07-19 \\ \hline
        Telescope 
        & BRC-250M                      & ChV-400         & ChV-400       & ChV-400 \\ \hline
        Exposure, s 
        & 30                            & 30              & 60            & 60\\ \hline
        $\hat{V}_{x}$, pix/fr 
        & 0.06                          & 0.72            & -0.06         & -0.11 \\ \hline
        $\hat{V}_{y}$, pix/fr 
        & 0.37                          & 0.51            & 0.58          & -0.21 \\ \hline
        $\hat{V}_{RA}$, "/fr 
        & -0.11                         & -1.09           & 0.07          & 0.19 \\ \hline
        $\hat{V}_{DE}$, "/fr 
        & -0.61                         & 0.76            & 1.34          & -0.32 \\ \hline
        $\hat{V}_{RAcat}$, "/fr 
        & 0.09                          & -1.06           & 0.13          & 0.14 \\ \hline
        $\hat{V}_{DEcat}$, "/fr 
        & -0.52                         & 0.88            & 0.83          & -0.28 \\ \hline
        $\hat{V}$, pix/fr 
        & \textbf{0.37}                  & \textbf{0.88}   & \textbf{0.59} & \textbf{0.24} \\ \hline
        $\hat{V}$, "/fr 
        & 0.62                          & 1.33            & 1.35          & 0.31 \\ \hline
        ${V}_{cat}$, "/fr 
        & 0.53                          & 1.38            & 0.84          & 0.38 \\ \hline
        Average FWHM, pix 
        & 3.35                          & 4.59            & 5.12          & 4.92 \\ \hline
        Average SNR, "/fr 
        & 10.31                         & 7.76            & 7.26          & 42.14\\ \hline
        $\hat{\sigma}_{0}$, pix (UCAC4) 
        & \textbf{0.38}                  & \textbf{0.39}   & \textbf{0.39} & \textbf{0.26} \\ \hline
        $\hat{\sigma}_{0}$, " 
        & 0.54                          & 0.62            & 0.57          & 0.36 \\ \hline
        ${Mag}_{cat}$, $^{m}$
        & 15.32                         & 15.91           & 16.56         & 13.71 \\ \hline
        Asteroid-Moon dist., deg
        & 67.5                          & 79.5            & 83.5          & 84 \\ \hline
        Moon phase \%
        & 99                             & 79              & 76            & 14 \\ \hline 
        $k=\hat{V} / \hat{\sigma}_{0}$ 
        & \textbf{0.97}                  & \textbf{2.26}   & \textbf{1.51} & \textbf{0.92} \\ \hline
\end{tabular}
\end{table}

\textbf{Discovery of the sungrazing comet C/2012 S1 (ISON).} On September 21, 2012 the sungrazing comet C/2012 S1 (ISON) was discovered (Fig. 6) at the ISON-Kislovodsk Observatory \citep[][]{kislovodsk} of the International Scientific Optical Network (ISON) project \citep[][]{molotov}, \citep[][]{mpc}. Information about observatory and telescope is available in the Table 1.
\begin{figure}
        \label{pic5}
        \centering
        \includegraphics[width=9cm]{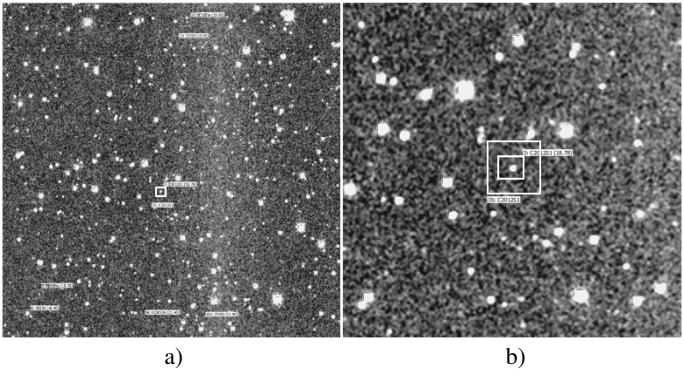}
        \small 
        \begin{tabular}{R{2.2cm} C{3.4cm} L{3cm}}
                a) & & b) 
        \end{tabular}
        \caption{Sungrazing comet C/2012 S1 (ISON) at the moment of discovery in the center of crop of CCD-frame with field of view 20 x 20 arcminutes (a), 8 x 8 arcminutes (b)}
\end{figure}
At the moment of discovery, the magnitude of the comet was equal to 18.8${^m}$, and its coma had 10 arc seconds in diameter that corresponds to 50 000 km at a heliocentric distance of 6.75 au. Its apparent motion velocity at the moment of discovery was equal to 0.8 pixels per frame. The size of the comet image in the frame was about five pixels. In Fig. 7a the cell size corresponds to the size of the pixel and is equal to 2~arc seconds. Within 26 minutes of the observation, the image of the comet has been moved by three pixels in the series of 4 CCD-frames (Fig. 7b). 

\begin{figure}[H]
        \label{pic6}
        \centering
        \includegraphics[width=9cm]{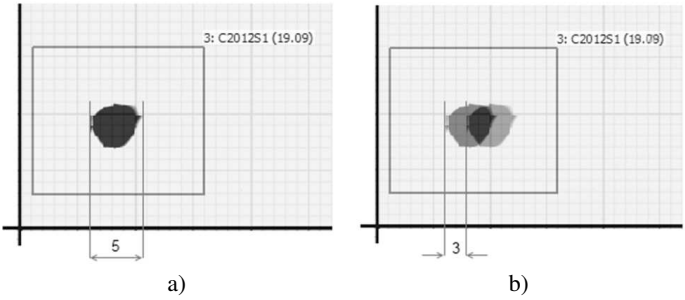}
        \small 
        \begin{tabular}{R{2.2cm} C{3.4cm} L{3cm}}
                a) & & b) 
        \end{tabular}
        \caption{a) Images of C/2012 S1 (ISON) comet on CCD-frames: the image size is five pixels (a), the shift of comet image between the first and the fourth CCD-frames of series is three pixels (b)}
\end{figure}

C/2012 S1 (ISON) comet (Fig. 8) was detected using the CoLiTec software for automated asteroids and comets discoveries \citep[][]{savanevych1} with the implemented method of detection using Fisher f-criterion (5).
\begin{figure}[H]
        \label{pic7}
        \centering
        \includegraphics[width=9cm]{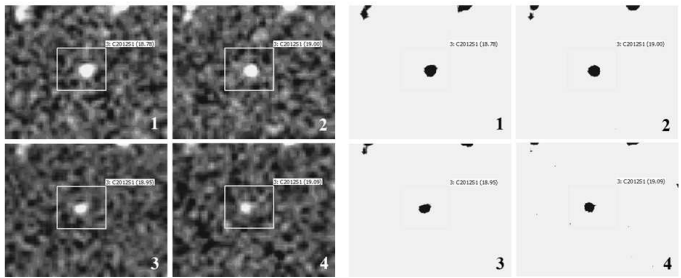}
        \caption{Sungrazing comet C/2012 S1 (ISON) in a series of four CCD-frames}
\end{figure}
C/2012 S1 (ISON) comet was disintegrated at an extremely small perihelion distance of about 1 million km on the day of perihelion passage, on November 28, 2013. Its disintegration was caused by the Sun's tidal forces and the significant mass loss due to the alterations in the moments of inertia of its nucleus. Despite having a short visible life time for our observations, this comet supplemented our knowledge of cometary astronomy.

\section{Conclusions}
We proposed a computational method for the detection of objects with the near-zero apparent motion on a series of CCD-frames, which is based on the Fisher f-criterion \citep[][]{phillips} instead of using the traditional decision rules that based on the maximum likelihood criterion \citep[][]{myung}.

For the analysis of the indicators of quality of the apparent motion detection, we applied statistical and in situ modeling methods and determined their conditional probabilities of true detection (CPTD) of the near-zero motion on a series of CCD-frames.

The statistical modeling showed that the most effective and adaptive method for the apparent motion detection is the substitutional method of maximum likelihood using the external estimation of RMS errors (C.13) (Fig. 1). But the process of determining the critical values of decisive statistics is very time- and resource-consuming in the rapidly changing observational conditions. By this reason, we recommended to apply the method of the near-zero apparent motion detection for the subclass of objects with $3\sigma$-velocity using Fisher f-criterion (5) for series with the number of frames $N_{fr} = 4$ or more (Fig. 1). The condition of a large number of frames in the series also makes the proposed method not inferior to other methods of apparent motion detection by CPTD.

When studying the indicators of quality of near-zero apparent motion detection by the in situ modeling method the objects from the internal catalog fixed on a series of CCD-frames were used as in situ data. It was found that in the case when the velocity does not exceed 3 RMS errors in object position per frame, the most effective method for near-zero apparent motion detection is the method which uses Fisher f-criterion (Fig. 3, 4). When compared with other methods, this method is stable at the equipment replacement (Fig. 5).

The proposed method for detection of the objects with $3\sigma$-velocity apparent motion using Fisher f-criterion was verified by authors and implemented in the embedded plugin developed in the CoLiTec software for automated discovery of asteroids and comets \citep[][]{savanevych1}. 

Among the other objects detected and discovered with this plugin, there was the sungrazing comet C/2012 S1 (ISON) \citep[][]{mpc}. The velocity of the comet apparent motion at the moment of discovery was equal to 0.8 pixels per CCD-frame. Image size of the comet on the frame was about five pixels (Fig. 7a). Within 26 minutes of the observation, the image of the comet had moved by three pixels in the series of four CCD-frames (Fig. 7b). So, it was considered to belong to the subclass of SSOs that have a velocity of apparent motion between CCD-frames not exceeding three RMS errors $\sigma$ of measurements of its position ($\varepsilonup=3\sigma$). In total, about 15\% of SSO objects with $\varepsilonup$-velocity apparent motion in the CCD-frames. These are the objects beyond the Jupiter's orbit as well as asteroids heading straight to the Earth. 

\section{Acknowledgments}
The authors thank observatories that have implemented CoLiTec software for observations. We especially thank Vitaly Nevski and Artyom Novichonok for their discovery of ISON comet and others SSOs. We are grateful to the reviewer for their helpful remarks that improved our paper and, in particular, for the suggestion "to add a few real-life examples, where the method provides a detection of motion for an object that would otherwise be difficult to detect". We express our gratitude to Mr. W. Thuillot, coordinator of the Gaia-FUN-SSO network \citep[][]{thuillot}, for the approval of CoLiTec as a well-adapted software to the Gaia-FUN-SSO conditions of observation (\url{https://gaiafunsso.imcce.fr}). 

Research is supported by the APVV-15-0458 grant and the VVGS-2016-72608 internal grant of the Faculty of Science, P. J. Safarik University in Kosice (Slovakia).

The CoLiTec software is available on \url{http://neoastrosoft.com}.

\begin{appendix}
\section{Model of the motion parameters}
The model of rectilinear and uniform motion of an object along each coordinate independently can be represented with the set of equations:
\begin{equation}\label{eq1}
        x_{k}(\theta_{x})=x_{0}+V_{x}(\tau_{k}-\tau_{0});
\end{equation}
\begin{equation}\label{eq2}
        y_{k}(\theta_{y})=y_{0}+V_{y}(\tau_{k}-\tau_{0}),
\end{equation}
where 
$k(i,n)=k$ is the index number of measurement in the set, namely, $i$-th measurement of $n_{fr}$-th CCD-frame with the observed object;

$x_{0}$, $y_{0}$ are the coordinates of object from the set of measurements at the time $\tau_{0}$ of the base frame timing;

$V_{x}$, $V_{y}$ are the apparent velocities of object along each coordinate:
\begin{equation}\label{eq3}
        \theta_{x}=(x_{0},V_{x})^T;
\end{equation}
\begin{equation}\label{eq4}
        \theta_{y}=(y_{0},V_{y})^T;
\end{equation}
are the vectors of the parameters of the apparent motion of the object along each coordinate, respectively.

The measured coordinates $x_{k}$, $y_{k}$ at the time $\tau_{k}$ are also determined by the parameters of the apparent motion of object in CCD-frame and can be calculated according to Equations (A.1) and (A.2).

So, the set of $N_{fr}$ measurements of $n_{fr}$-th frame timing at the time $\tau_{n}$ is generated from observations of a certain area of the celestial sphere. One frame of the series is a base CCD-frame, and time of its anchoring is the base frame timing $\tau_{0}$. The asteroid image on $n_{fr}$-th frame has no differences from the images of stars on the same frame.
Results of intra-frame processing (one object per CCD-frame) can be presented as the $Y_{in}$ measurement ($i$-th measurement on the $n_{fr}$-th frame). In general, the $i$-th measurement on the $n_{fr}$-th frame contains estimates of coordinates $Y_{Kin}=\{x_{in};y_{in}\}$ and brightness $A_{in}$ of the object: $Y_{in}=\{Y_{Kin};A_{in}\}$. We used a rectangular coordinate system (CS) with the center located in the upper left corner of CCD-frame. It is assumed that all the positional measurements of the object are previously transformed into coordinate system of the base CCD-frame.

A set of measurements (no more than one in the frame), belonging to the object, has the form as follows:
\begin{equation}\label{eq5}
        \Omega_{set}=(Y_{K1(i,1)},...,Y_{Kk(i,n)},...,Y_{KNmea(i,Nfr)})=\nonumber
\end{equation}
\begin{equation}\label{eq5}
        =((x_{1},y_{1}),...,(x_{k},y_{k}),...,(x_{Nmea},y_{Nmea})),
\end{equation}
where $N_{mea}$ is the number of the position measurements of the object in $N_{fr}$ frames.
Measurements $Y_{k}$ from the set $\Omega_{set}$ (A.5) of measurements are selected by the rule of no more than one measurement per frame. Measurements of the object positions can not be obtained in all CCD-frames.
Therefore, the number of measurements which belong to the object in certain set of measurements will generally be equal to $N_{mea}$ ($N_{mea} \leq N_{fr}$).

It is supposed that the observational conditions are practically unchanged during observations of object with near-zero apparent motion. So, the RMS errors of estimates of its coordinates in the different CCD-frames are almost identical. Deviations of estimates of coordinates of this object, which belong to the same set $\Omega_{set}$ of measurements, are independent of each other both inside the one measurement and between measurements obtained in different frames. Deviations of coordinates are normally distributed \citep[][]{kuzmyn}, have a zero mathematical expectation and unknown variances (standard deviations) $\sigma_{x}^2$, $\sigma_{y}^2$.

\section{Likelihood function for detection of a near-zero apparent motion}
This common density distribution for $H_{0}$ hypothesis (1), assuming that the object is a star with zero rate apparent motion, is defined as follows:
\begin{equation}
\label{eq8}
f_{0}(\bar{x}, \bar{y}, \sigma)=\prod\limits_{k=1}^{N_{mea}}[N_{xk}(\bar{x},\sigma^2)N_{yk}(\bar{y},\sigma^2)],
\end{equation}
where $\bar{x}$, $\bar{y}$ are the coordinates of the object;

$N_{z}(m_{z},\sigma^{2})=\frac{1}{\sqrt{2\pi\sigma}}exp(-\frac{1}{2\sigma^{2}}(z-m_{z})^{2})$ is the density of normal distribution with mathematical expectation $m_{z}$ and variance $\sigma^2$ in $z$ point.

The common density distribution for $H_{1}$ hypothesis (2) is defined otherwise. Namely, the coordinates $x_{k}(\theta_{x})$, $y_{k}(\theta_{y})$ at the time $\tau_{k}$, calculated from Equations (A.1) and (A.2), must be used instead of the object's position parameters $\bar{x}$, $\bar{y}$ :
\begin{equation}
\label{eq9}
f_{1}(\theta, \sigma)=\prod\limits_{k=1}^{N_{mea}}[N_{xk}(x_{k}(\theta_{x}),\sigma^2)N_{yk}(y_{k}(\theta_{y}),\sigma^2)].
\end{equation}

Absence of information on the position of the object, its apparent motion and variance of estimates of object position in a set of measurements leads to the necessity of using the substitutional decision rule \citep[][]{lehman,morey}. In this case, the statistics for distinguishing these hypotheses is the LR estimate $\hat{\lambda}(\Omega_{set})$ \citep[][]{morey}.

\section{Evaluation of parameters for substitutional methods of maximum likelihood detection of a near-zero apparent motion}
OLS-evaluation of the parameters of the object's apparent motion may be represented in the scalar form \citep[][]{kuzmyn}:
\begin{equation}
\label{eq10}    
\hat{x}_{0}=\frac{DA_{x}-CB_{x}}{N_{mea}D-C^2};\hat{V}_{x}=\frac{N_{mea}B_{x}-CA_{x}}{N_{mea}D-C^2};
\end{equation}
\begin{equation}
\label{eq11}    
\hat{y}_{0}=\frac{DA_{y}-CB_{y}}{N_{mea}D-C^2};\hat{V}_{y}=\frac{N_{mea}B_{y}-CA_{y}}{N_{mea}D-C^2},
\end{equation}
where $A_{x}=\sum\limits_{k=1}^{N_{mea}}x_{k}$; $A_{y}=\sum\limits_{k=1}^{N_{mea}}y_{k}$; $B_{x}=\sum\limits_{k=1}^{N_{mea}}\Delta_{\tau k}x_{k}$; $B_{y}=\sum\limits_{k=1}^{N_{mea}}\Delta_{\tau k}y_{k}$; $C=\sum\limits_{k=1}^{N_{mea}}\Delta_{\tau k}$; $D=\sum\limits_{k=1}^{N_{mea}}\Delta_{\tau k}^2$;

$\Delta_{\tau k}=(\tau_{k}-\tau_{0})$ is the difference between the time $\tau_{0}$ of the base frame and time $\tau_{k}$ of the frame, in which the $k$-th measurement is obtained.

The interpolated coordinates of the object in the $k$-th frame are represented as
\begin{equation}
\label{eq12}    
\hat{x}_{k}=\hat{x}_{k}(\hat{\theta}_{x})=\hat{x}_{0}(\hat{\theta}_{x})+\hat{V}_{x}(\hat{\theta}_{x})(\tau_{k}-\tau_{0});
\end{equation}
\begin{equation}
\label{eq13}    
\hat{y}_{k}=\hat{y}_{k}(\hat{\theta}_{y})=\hat{y}_{0}(\hat{\theta}_{y})+\hat{V}_{y}(\hat{\theta}_{y})(\tau_{k}-\tau_{0}).
\end{equation}

Thus, for each ($k$-th) measurement from $N_{mea}$ measurements of the set $\Omega_{set}$ (A.5), we have:
\begin{itemize}
        \item the unknown real position of the object $x_{k}(\theta_{x})$, $y_{k}(\theta_{y})$;
        \item the measured object coordinates $x_{k}$, $y_{k}$ at the time $\tau_{k}$ in the coordinate system of the base frame;
        \item the interpolated coordinates $(\hat{x}_{k}, \hat{y}_{k}) = \hat{x}_{k}(\hat{\theta}_{x})$, $\hat{y}_{k}(\hat{\theta}_{y})$ defined by Equations (C.3) and (C.4).
\end{itemize}

\textbf{The variance of the object's positional estimates in a set of measurements.} Using the measured $x_{k}$, $y_{k}$ (A.1), (A.2) and the interpolated $(\hat{x}_{k}$, $\hat{y}_{k})$ (C.3), (C.4) coordinates, the variance estimates $\hat{\sigma}_{x}^2$ and $\hat{\sigma}_{y}^2$ (hereinafter - variances) of the object's positions can be represented as:
\begin{equation}
\label{eq14}
\hat{\sigma}_{x}^2 = \sum\limits_{k=1}^{N_{mea}}(x_{k} - \hat{x}_{k}(\hat{\theta}_{x}))^2 / (N_{mea} - m);
\end{equation}
\begin{equation}
\label{eq15}
\hat{\sigma}_{y}^2 = \sum\limits_{k=1}^{N_{mea}}(y_{k} - \hat{y}_{k}(\hat{\theta}_{y}))^2 / (N_{mea} - m),
\end{equation}
where $m=2$ is the number of parameters of the apparent motion along each coordinate in a set of measurements.

Assuming the validity of the hypothesis about zero ($H_{0}$) and near-zero ($H_{1}$) apparent motions, the conditional variances $\hat{\sigma}_{0}^2$, $\hat{\sigma}_{1}^2$ of the object's position can be represented as:
\begin{equation}
\label{eq16}
\hat{\sigma}_{0}^2 = \frac{R_{0}^2}{2(N_{mea}-m)};
\end{equation}
\begin{equation}
\label{eq17}
\hat{\sigma}_{1}^2 = \frac{R_{1}^2}{2(N_{mea}-m)},
\end{equation}
where 
\begin{equation}
\label{eq18}
R_{0}^2 = \sum\limits_{k=1}^{N_{mea}}((x_{k} - \hat{\bar{x}})^2 + (y_{k} - \hat{\bar{y}})^2);
\end{equation}
\begin{equation}
\label{eq19}
R_{1}^2 = \sum\limits_{k=1}^{N_{mea}}((x_{k} - \hat{x}_{k}(\hat{\theta}_{x}))^2 + (y_{k} - \hat{y}_{k}(\hat{\theta}_{y}))^2),
\end{equation}
are the residual sums of the squared deviations of object's positions \citep[][]{burden}.

We note also that the variance of the positions in a set of measurements can be obtained by the external data, for example, from measurements of another objects on a series of CCD-frames. Hence, the required estimate is a variance estimation of all position measurements of objects detected in CCD-frame and identified in any astrometric catalog.

\textbf{Substitutional methods for maximum likelihood detection of a near-zero apparent motion} may operate with unknown real position $x_{k}(\theta_{x})$, $y_{k}(\theta_{y})$ of the object at a time $\tau_{k}$ and unknown variances $\sigma_{x}^2$, $\sigma_{y}^2$ of the object's position in CCD-frames.

It is easy to show that in the latter case the substitutional method can be represented as
\begin{equation}
\label{eq20}
\frac{R_{0}^2 - R_{1}^2}{R_{0}^2 R_{1}^2} \geq \frac{\ln(\lambda_{cr})}{A N_{mea}},
\end{equation}
where $\lambda_{cr}$ is the maximum allowable (critical) value of the LR estimate for the detection of a near-zero apparent motion; $A = 2(N_{mea} - m)$.

If the variance $\sigma^2$ of the object's position is known, the substitutional method can be represented as
\begin{equation}
\label{eq21}
R_{0}^2 - R_{1}^2 \geq 2\sigma^2 \ln(\lambda_{cr}),
\end{equation}

In that case, if the external variance estimation $\hat{\sigma}_{out}^2$ of the position is used, the substitutional method takes the form: 
\begin{equation}
\label{eq22}
\frac{R_{0}^2 - R_{1}^2}{\hat{\sigma}_{out}^2} \geq 2\ln(\lambda_{cr}),
\end{equation}

\end{appendix}
\end{document}